\documentclass[journal=jpcafh,manuscript=article]{achemso}

\usepackage[version=3]{mhchem} 
\usepackage[T1]{fontenc}       

\newcommand{\rr}{{\bf{r}}}

\newcommand{\curl}{\nabla\times}
\newcommand{\mi}{{\mathrm{i}}}

\usepackage{color}

\usepackage{subfigure}

\author{Johannes Fiedler}
\email{johannes.fiedler@physik.uni-freiburg.de}
\affiliation{Physikalisches Institut, Albert-Ludwigs-Universit{\"a}t Freiburg, Hermann-Herder-Str. 3, 79104 Freiburg, Germany}

\author{Priyadarshini Thiyam}
\email{thiyam@kth.se}
\affiliation{Department of Materials Science and Engineering, KTH, Royal Institute of Technology, SE-100 44 Stockholm, Sweden}
\alsoaffiliation{Centre for Materials Science and Nanotechnology,
Department of Physics, University of Oslo, P. O. Box 1048 Blindern,
NO-0316 Oslo, Norway}

\author{Anurag Kurumbail}
\affiliation{Department of Energy and Process Engineering, Norwegian University of Science and Technology, NO-7491 Trondheim, Norway}

\author{Friedrich Burger}
\affiliation{Physikalisches Institut, Albert-Ludwigs-Universit{\"a}t Freiburg, Hermann-Herder-Str. 3, 79104 Freiburg, Germany}

\author{Michael Walter}
\affiliation{Physikalisches Institut, Albert-Ludwigs-Universit{\"a}t Freiburg, Hermann-Herder-Str. 3, 79104 Freiburg, Germany}
\alsoaffiliation{FIT Freiburg Centre for Interactive Materials and Bioinspired Technologies, Georges-K\"ohler-Allee 105, 79110 Freiburg, Germany}
\alsoaffiliation{Fraunhofer IWM, W\"ohlerstrasse 11, D-79108 Freiburg i. Br., Germany}

\author{Clas Persson}
\affiliation{Department of Materials Science and Engineering, KTH, Royal Institute of Technology, SE-100 44 Stockholm, Sweden}
\alsoaffiliation{Centre for Materials Science and Nanotechnology,
Department of Physics, University of Oslo, P. O. Box 1048 Blindern,
NO-0316 Oslo, Norway}

\author{Iver Brevik}
\affiliation{Department of Energy and Process Engineering, Norwegian University of Science and Technology, NO-7491 Trondheim, Norway}

\author{Drew F. Parsons}
\email{d.parsons@murdoch.edu.au}
\affiliation{School of Engineering and IT, Murdoch University, 90 South St, Murdoch, WA 6150, Australia}

\author{Mathias  Bostr{\"o}m}
\email{mathias.a.bostrom@ntnu.no}
\affiliation{Department of Energy and Process Engineering, Norwegian University of Science and Technology, NO-7491 Trondheim, Norway}

\author{Stefan Y. Buhmann}
\affiliation{Physikalisches Institut, Albert-Ludwigs-Universit{\"a}t Freiburg, Hermann-Herder-Str. 3, 79104 Freiburg, Germany}

\alsoaffiliation{Freiburg Institute for Advanced Studies, Albert-Ludwigs-Universit{\"a}t Freiburg, Albertstr. 19, 79104 Freiburg, Germany}
\email{stefan.buhmann@physik.uni-freiburg.de}

\title[Effective Polarisability Models]
  {Effective Polarisability Models}

\abbreviations{IR,NMR,UV}

\begin{document}

\begin{tocentry}

\includegraphics[width=9cm]{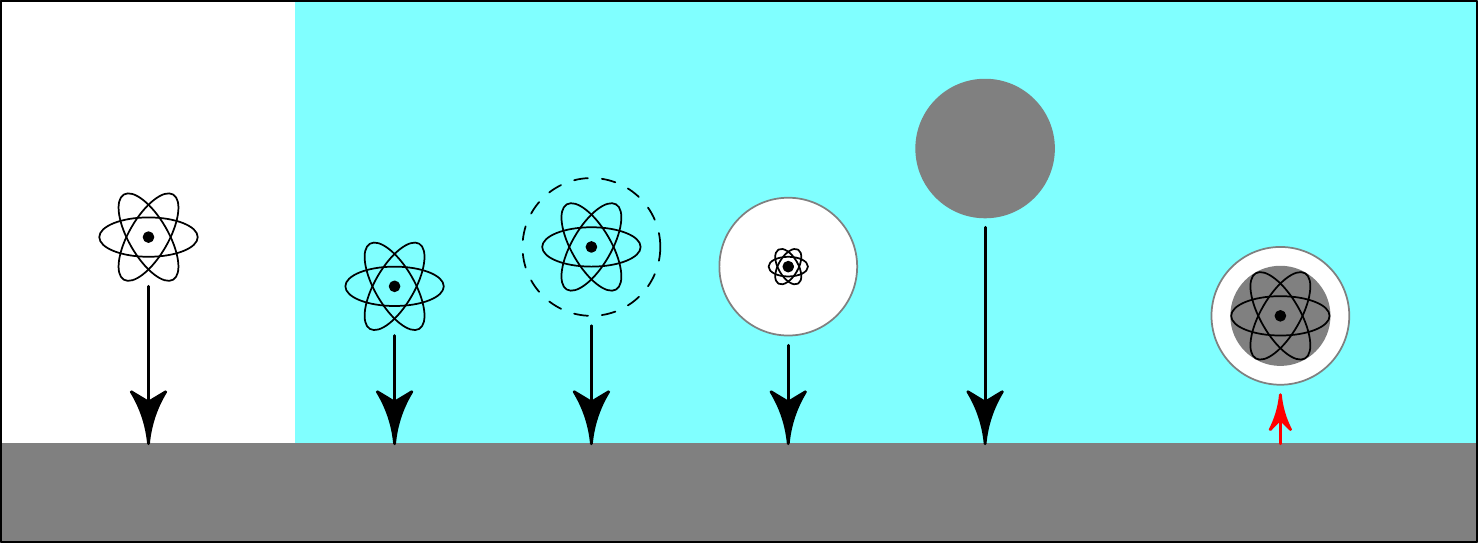}

\end{tocentry}

\begin{abstract}
Theories for the effective polarisability of a small particle in a medium are presented using different levels of approximation: we consider the virtual cavity, real cavity and the hard-sphere models as well as a continuous interpolation of the latter two. We present the respective hard-sphere and cavity radii as obtained from density-functional simulations as well as the resulting effective polarisabilities at discrete Matsubara frequencies. This enables us to account for macroscopic media in van der Waals interactions between  molecules in water and their Casimir--Polder interaction with an interface. 
\end{abstract}

\section{Introduction}
The optical behaviour of small object dissolved in a medium are of interest for a large number of investigations, such as optofluids\cite{Roadmap17}, medium-assisted density-functional theory (DFT)\cite{Held14}, nanomedicine\cite{doi:10.2217/nnm.13.99}, hydrogen storage\cite{doi:10.1021/nl071436g}, bio-organics\cite{Cheng2015} and photodynamic therapy\cite{B711141J}. The impact of a cavity around the particles is large compared to the typically studied situations where the particles are considered without an environment.

In this  fundamental study we derive different models for the effective polarisability of small particles in a medium. The level of accuracy for six different approximations will be discussed and our results exploited to calculate Casimir--Polder forces in a media. The new theoretical results are then applied to greenhouse molecules \cite{Thiyam2015316}, and other gas molecules, dissolved in water.  After water vapour and carbon dioxide, methane and nitrous oxide are the most important long-lived greenhouse gases in the atmosphere.  Release of greenhouse gases from surfaces in water is influenced by the Casimir--Polder interaction. In order to estimate the Casimir-Polder binding to surfaces one needs to have accurate effective polarisabilities of the greenhouse gas molecules in water. As input parameters in our models we require the frequency dependent polarisability of the molecules in vacuum, hard sphere radii, cavity radii in water, and the dielectric function of water. We use ab initio quantum chemical calculations to calculate accurately the frequency dependent polarisability of different gas molecules in vacuum. The result is given as a set of parameterised functions which enables easy access in future investigations that require these molecular polarisabilities.
We also provide the theory for determining the molecular radii  and cavity radii of different greenhouse-gas molecules in water.

These parameterised functions and tabulated data for radii, together with the new theories and tabulated  dielectric function of water for discrete Matsubara frequencies, enable us to calculate the effective polarisability of gas molecules in water. This is a key quantity needed in future studies of binding or release of  gas molecules near solid-water interfaces. As an illustrative example we consider non-retarded van--der--Waals potential of two molecules in water and their Casimir--Polder potential near a perfect metal surface. This enables us to compare how sensitive the results are to  different approximations done when evaluating the effective polarisability.  It turns out that the theoretical sensitivity can be quite large; even the sign of the force on molecules close to a metal surface can change if one goes from one effective polarisability model to another. There is thus a need for experimental information to identify the best model.

\section{Free-space polarisability of gas molecules}

The Casimir--Polder energy is determined from the effective polarisability of a molecule in a medium at imaginary frequencies $\mi\xi$. Each of the models of effective polarisability considered in this manuscript describes a different representation of the transformation induced by the medium on the underlying dynamic polarisability $\alpha(\mi \xi_n)$ of the given  molecule in vacuum (the free space polarisability).  We have computed the free space polarisability of each molecule by quantum chemical methods. The closed shell gas molecules,  $\mathrm{CH}_4$, $\mathrm{CO}_2$, $\mathrm{N_2O}$, $\mathrm{O_3}$, $\mathrm{N_2}$, $\mathrm{CO}$, and $\mathrm{H_2S}$ were calculated at the Coupled Cluster Singles and Doubles (CCSD) level of theory \cite{HampelPetersonWerner1992} using Molpro \cite{MOLPRO2012}. The open shell (paramagnetic) molecules $\mathrm{O_2}$ and $\mathrm{NO_2}$ were calculated using Turbomole \cite{turbomole64} at a Density Functional Theory (DFT) level with a hybrid PBE0 functional \cite{AdamoBarone1999}. Ground states involved higher-spin states, $\mathrm{O_2}$ having spin multiplicity 3 (two unpaired electrons), $\mathrm{NO_2}$ with multiplicity 2 (one unpaired electron). For both CCSD and DFT/PBE0 calculations we employed an augmented correlation-consistent basis set, aug-cc-pVQZ \cite{PetersonDunning2002}. The geometry of each molecule was optimised by energy minimisation before evaluating its polarisability.  The full anisotropic polarisability tensor was evaluated and anisotropic effects are known to have an impact on Casimir-Polder interactions \cite{ParsonsDenizNinham2009,ThiyamParasharShajeshPerssonSchadenBrevikParsonsMiltonMalyiBostrom2015}. However for simplicity  we use the isotropic average, $\alpha=(\alpha_{xx}+\alpha_{yy}+\alpha_{zz})/3$, in this manuscript.

Quantum chemical calculations of dynamic polarisabilities were performed at the Matsubara frequencies $\mi\xi_n = \mi2\pi k_B Tn/\hbar$ for $T=298.15$ K with $n=0,1\dots,2100$. It is convenient to represent the polarisability at an arbitrary imaginary frequency $i\xi$ by fitting to an oscillator model
\begin{equation}
\alpha(i \xi)=\sum_j\frac{\alpha_j}{1+(\xi/\omega_j)^2}
\label{Eq2}
\end{equation}
A 5-mode fit has previously been found to describe the dynamic polarisability accurately to a 0.02\% relative error \cite{ParsonsNinham2010dynpol}. The adjusted parameters for a 5-mode model fitted to agree with the free space polarisability obtained from ab initio calculations are given in Table 1.

\begin{table}
\caption{The weights ($\alpha_j\, [10^{-42} \, \mathrm{A^2 s^4 kg^{-1}}]$ in SI-units which transform to CGS units via $[\alpha_{SI}] = 4 \pi\varepsilon_0 10^{-30} \text{\AA}^3$) and characteristic frequencies ($\omega_j \,[10^{16} \, \mathrm{rad \, s^{-1}}]$ in SI-units, which transform to CGS units via $[\omega_{SI}] = 2\pi e/\hbar \, \mathrm{eV}$) for five-mode
  London fits of the dynamic polarisabilities of four greenhouse gas molecules CH$_4$, CO$_2$, N$_2$O, O$_3$, and other atmospheric gas molecules.
}

  \begin{tabular}{l|r|l||r|l||r|l||r|l||r|l}
  &  \multicolumn{2}{c||}{mode 1}  &  \multicolumn{2}{c||}{mode 2}  &  \multicolumn{2}{c||}{mode 3}  &  \multicolumn{2}{c||}{mode 4}  &  \multicolumn{2}{c}{mode 5}     \\
&   \multicolumn{1}{c|}{   $\alpha_1$} 
& \multicolumn{1}{c||}{$\omega_1$ }
&  \multicolumn{1}{c|}{$\alpha_2$}
&  \multicolumn{1}{c||}{$\omega_2$}
&  \multicolumn{1}{c|}{$\alpha_3$}
& \multicolumn{1}{c||}{$\omega_3$} 
&\multicolumn{1}{c|}{$\alpha_4$} 
& \multicolumn{1}{c||}{$\omega_4$}
&  \multicolumn{1}{c|}{$\alpha_5$} 
&  \multicolumn{1}{c}{$\omega_5$} 
\\ \hline
$\mathrm{CH}_4$ &89.3  & 1.75 & 137.5 & 2.56 & 41.2  & 4.42 & 2.78 & 10   & 0.18  & 48.3 \\
$\mathrm{CO}_2$ & 131.7& 1.95 & 116   & 3.14 & 41.5  & 6.2  & 6.43 & 13.3 & 0.43  & 51.5  \\
$\mathrm{N_2O}$ & 121.5& 1.64 & 146.2 & 54.9 & 54.9  & 5.59 & 8.05 & 12.3 & 0.41  & 53.9 \\
$\mathrm{O_3}$  & 72   & 0.89 & 130   & 2.28 & 98.7  & 4.32 & 21   & 9.82 & 0.89  & 36.5 \\
$\mathrm{O_2}$  & 38   & 1.37 & 85.4  & 2.78 & 41.1  & 5.42 & 8.16 & 10.9 & 0.76  & 29.5 \\
$\mathrm{N_2}$  & 90   & 2.14 & 10    & 3.27 & 29.9  & 5.92 & 3.55 & 13.2 & 0.23  & 60.5 \\
$\mathrm{CO}$   & 49.1 & 1.43 & 120   & 2.46 & 41.7  & 4.95 & 6.89 & 11   & 0.39  & 45.7 \\
$\mathrm{NO_2}$ & 16.5 & 0.63 & 124   & 1.65 & 134.1 & 3.57 & 28.8 & 8.39 & 1.19  & 31.6 \\
$\mathrm{H_2S}$ & 55.1 & 1.12 & 99    & 3.25 & 251.6 & 1.86 & 2.33 & 9.58 & 0.98  & 34.2 \\
  \end{tabular}

\end{table}

\section{Casimir--Polder and van der Waals potential}
The Casimir--Polder force arises when a small particle comes close to a dielectric surface \cite{Casimir1948}. It is caused by the fluctuations of the ground-state electromagnetic field \cite{Buhmann2012}. Performing the field quantisation and applying second-order perturbation theory to the dipole-electric field Hamiltonian $\hat{H}=-\hat{\bf{d}}\cdot\hat{\bf{E}}$ for the ground state of field and particle, one finds the Casimir--Polder potential acting on a particle located at $\rr_A$ for $T=0$ \cite{Buhmann2012}
\begin{equation}
 U_{CP}(\rr_A) = \frac{\hbar \mu_0}{2\pi} \int\limits_0^\infty \mathrm d \xi \, \xi^2 {\alpha}(\mi\xi)\operatorname{tr}\left[ {\bf{G}}^{(S)}(\rr_A,\rr_A,\mi\xi)\right] \, , \label{eq:UCP}
\end{equation}
with the reduced Planck constant $\hbar$, the vacuum permeability $\mu_0$, the particle polarisability $\boldsymbol{\alpha}(\mi\xi)$ at imaginary frequencies $\mi\xi$ and the scattering part of the dyadic Green's function ${\bf{G}}^{(S)}$, that satisfies the vector Helmholtz equation
\begin{equation}
 \curl  \curl{\bf{G}}({\rr,\rr'},\omega) -  \frac{\omega^2}{c^2}\varepsilon({\bf{r}},\omega){\bf{G}}({\rr,\rr'},\omega) = \boldsymbol\delta(\rr-\rr') \, , \label{eq:GREEN}
\end{equation}
with the relative permittivity $\varepsilon(\rr,\omega)$. 
The dyadic Green function can be separated into the free propagation through the bulk medium ${\bf{G}}^{(0)}$ and the scattering part ${\bf{G}}^{(S)}$, ${\bf{G}}={\bf{G}}^{(0)}+{\bf{G}}^{(S)}$. The Casimir--Polder potential, Eq.~(\ref{eq:UCP}), shall be understood due to a virtual photon with frequency $\mi\xi$ being created by the particle at $\rr_A$ and back scattered at the dielectric surface as expressed by the scattering Green function. The strength of its fraction to the Casimir--Polder potential is weighted by the polarisability of the particle. Because the fluctuations of the fields ensue at all frequencies, a superposition of all scattering processes results in the Casimir--Polder potential. 

Assuming that the particle is located in the non-retarded regime in front of planar surface the Casimir--Polder potential reduces to the established $C_3$-potential. This situation is described by the scattering Green function for a planar two-layer system\cite{Fiedler15}
\begin{equation}
 {\bf{G}}(\rr,\rr,\omega) = \frac{c^2}{32\pi\omega^2\varepsilon(\omega) z^3}\frac{\varepsilon_H(\omega)-\varepsilon(\omega)}{\varepsilon_H(\omega)+\varepsilon(\omega)}\operatorname{diag}(1,1,2) \, ,
\end{equation}
with the dielectric function $\varepsilon_H$ describing the electric response of the half-space. A perfectly conducting plate requires the limit $\varepsilon_H(\omega)\to\infty$. Due to the separation of  spatial and frequency dependencies of the scattering Green function, the Casimir--Polder potential, Eq.~(\ref{eq:UCP}), simplifies to the well-known result
\begin{equation}
 U_{CP}(z) = -\frac{C_3}{z^3} \, ,\qquad C_3 = \frac{\hbar }{16\pi^2\varepsilon_0} \int\limits_0^\infty \mathrm d \xi \frac{\alpha(\mi\xi)}{\varepsilon(\mi\xi)} \, . \label{eq:C3pure}
\end{equation}

In analogy, the van der Waals potential describing the interaction between two neutral, but polarisable particles can be derived via the fourth order perturbation of the two particle dipole-electric field Hamiltonian $\hat{H} = -\hat{\bf{d}}_A\cdot \hat{\bf{E}} - \hat{\bf{d}}_B\cdot\hat{\bf{E}}$, that will be performed for ground state fields and particles. For $T=0$ it results in \cite{Buhmann2012}
\begin{equation}
 U_{vdW}({\bf{r}}_A,{\bf{r}}_B) = -\frac{\hbar \mu_0^2}{2\pi}\int\limits_0^\infty \mathrm d \xi \, \xi^4 {\alpha}_A(\mi\xi){\alpha}_B(\mi\xi)\operatorname{tr}\left[ {\bf{G}}({\bf{r}}_A,{\bf{r}}_B,\mi\xi)\cdot{\bf{G}}({\bf{r}}_B,{\bf{r}}_A,\mi\xi)\right] \,,
\end{equation}
which has to be read from right to left and is due to a virtual photon which is created at particle $A$, propagates to particle $B$, where it interacts with its polarisability and is back-scattered to particle $A$. Again, the sum (integral) over all photon exchanges results in the van der Waals potential. 
According the non-retarded Casimir--Polder potential, the non-retarded Green function\cite{Buhmann2012}
\begin{equation}
{\bf{G}}(\rr_A,\rr_B,\omega) = -\frac{\mathrm e^{\mi k(\omega)\varrho}}{4\pi k^2(\omega) \varrho^3}\left\lbrace\left[1-\mi k(\omega)\varrho-k^2(\omega)\varrho^2\right]{\bf{I}}-\left[3-3\mi k(\omega)\varrho-k^2(\omega)\varrho^2\right]{\bf{e}}_\varrho\otimes{\bf{e}}_\varrho\right\rbrace \, ,
\end{equation}
with $\boldsymbol{\varrho}=\rr_B-\rr_A$, $\varrho=\left|\boldsymbol{\varrho}\right|$, ${\bf{e}}_\varrho=\boldsymbol{\varrho}/\varrho$ and $k^2(\omega)=\varepsilon(\omega)\omega^2/c^2$, for the propagation through a bulk medium can be applied. This results in
\begin{equation}
 U_{vdW}(r)=-\frac{C_6}{r^6} \, ,\qquad C_6 = \frac{3\hbar}{16\pi^3\varepsilon_0^2}\int\limits_0^\infty\mathrm d\xi \, \frac{\alpha_A(\mi\xi)\alpha_B(\mi\xi)}{\varepsilon^2(\mi\xi)} \, , 
\end{equation}
where $r$ denotes the distance between particle $A$ and $B$.

At finite temperature the transition dipole states are thermally distributed and follows Bose--Einstein statistics. \cite{Buhmann2012b} Performing contour integral technique, their poles for imaginary frequency arguments have to be taken into account which yields the Matsubara frequencies. 
Hence, the 
integral in Eq.~(\ref{eq:UCP}) turns into a discrete sum over these frequencies
\begin{equation}
 \hbar\int\limits_0^\infty\mathrm d \xi f(\xi) \rightarrow 2\pi k_B T \sum_{n=0}^\infty{}' f(\xi_n) = 2\pi k_B T \left[\frac{1}{2} f(\xi_0) +\sum_{n=1}^\infty f(\xi_n)\right]\, ,
\end{equation}
where the primed sum means that the zeroth term has to be weighted by $1/2$. Considering particles located in water in front of an infinite and perfectly conducting half plane, Eq.~(\ref{eq:C3pure}) can be applied to define the $C_3$ coefficient as
\begin{equation}
  C_3={\frac {k_B T} {8\pi^2\varepsilon_0}} \sum_{n=0}^{\infty}{}' {\frac {\alpha(\mi \xi_n)} {\varepsilon(\mi \xi_n)}} \,.
  \label{eq:C3}
\end{equation}
Two particles embedded in water results in $z^{-6}$ distance law. The $C_6$ coefficient reads
\begin{equation}
  C_6  = \frac{3 k_B T}{8\pi^2\varepsilon_0^2}\sum_{n=0}^\infty {}' \frac{\alpha_1(\mi \xi_n)\alpha_2(\mi\xi_n)}{\varepsilon^2(\mi\xi_n)} \, .
  \label{eq:C6}
\end{equation}

\begin{figure}
 \centering
 \subfigure[Sketch of Onsager's real cavity model. A particle (red dot) at position ${\bf{r}}_A$ is embedded in a medium with dielectricity $\varepsilon(\omega)$ (grey area) surrounded by a spherical vacuum cavity (white area). The scattering process from an external point ${\bf{r}}$ (green dot) is separated into the propagation to and back from the particle.]{\includegraphics[width=0.47\textwidth]{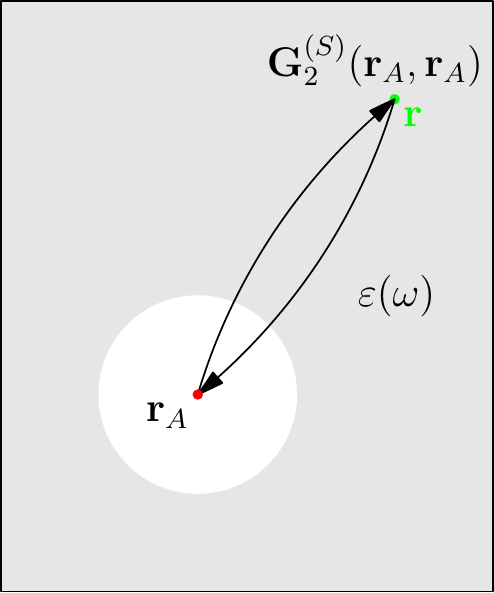}\label{fig:onsager}}\hfill
\subfigure[Sketch of Onsager's real cavity model for finite size particles. A spherical particle with radius $R$ and the dielectric function $\varepsilon_s$ is embedded in a medium with $\varepsilon(\omega)$ (grey area) surrounded by a spherical vacuum cavity (white area) with radius $R_C$. The scattering process from an external point ${\bf{r}}'$ (red dot) to another point ${\bf{r}}$ (green dot).]{\includegraphics[width=0.47\textwidth]{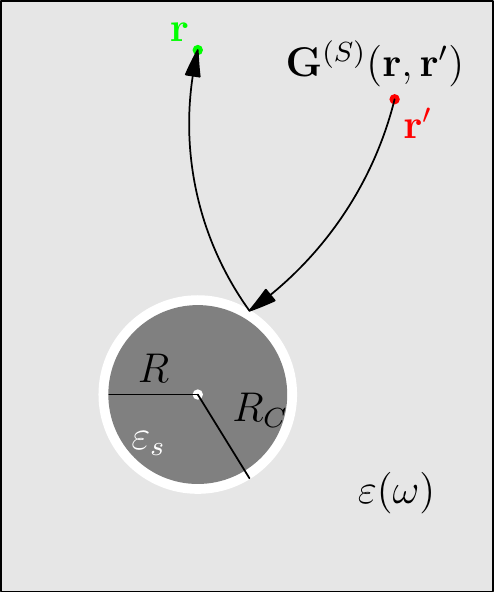}\label{fig:3layer}}
\caption{Sketch of the arrangements.}
\end{figure}

The advantage of these formulas is the separation into particle properties ${\alpha}(\mi\xi)$ and the scattering processes ${\bf{G}}^{(S)}(\rr_A,\rr_A,\mi\xi)$.  In order to describe the Casimir--Polder interaction in media, the scattering Green function changes its properties with respect to the dielectric function of the medium and to the geometrical shape of the cavity around the particle which rises additional resonances due to the cavity modes. Two different pictures for the consideration of the scattering processes will be used  describing the influence of a cavity effectively. In the first case, the particle is centred in the cavity and the scattering through the cavity's boundary is considered, see Fig.~\ref{fig:onsager}. Assuming that the scattering processes between the surface and the particle-cavity system are dominated by a single scattering effect (negligence of multiple scattering effects), the Green function factorises into a bulk propagation ${\bf{G}}$ (here: in water) and an effective transmission coefficient $R(\omega)$, which has to be found, through the cavity's surface, which includes the properties of the cavity and yields \cite{Sambale07}
\begin{equation}
 {\bf{G}}_{cav}(\rr,\rr',\omega) = R(\omega) {\bf{G}}(\rr,\rr',\omega) \, .
\end{equation}
Due to the linearity of the Casimir--Polder potential, Eq.~(\ref{eq:UCP}), in the polarisability and the scattering Green function, an excess polarisability can be defined as
\begin{equation}
{\alpha}^\star (\mi\xi) = R(\mi\xi) {\alpha}(\mi\xi) \, .
\end{equation}

In the second picture, we reverse the optical paths and describe the scattering effect as an effective reflection at a sphere which means that the corresponding initial and final points are located outside the sphere, see Fig.~\ref{fig:3layer}. The propagation is again represented by the free propagation in a bulk medium. Thus, we can write
\begin{equation}
 {\bf{G}}_{cav}({\bf{r}},{\bf{r}}',\omega) = {\bf{G}}({\bf{r}},{\bf{R}},\omega)\cdot{\alpha}^\star(\omega)\cdot {\bf{G}}({\bf{R}},{\bf{r}}',\omega) \,,\label{eq:excesspol}
\end{equation}
where ${\bf{R}}$ is a point at the spherical cavity's surface and the excess polarisability $\boldsymbol{\alpha}(\omega)$. In the following we analyse different models for such excess polarisabilities. In order to apply Eqs.~(\ref{eq:C3}) and (\ref{eq:C6}) the free-space polarisability have to be exchanged by the excess polarisability \cite{Burger17}
\begin{equation}
 \alpha(\mi\xi) \mapsto \alpha^\star(\mi\xi) \, .
\end{equation}

\section{Polarisability of particles embedded in medium}
In the following we introduce the established models estimating the effective electric polarisability of a particle embedded in a medium. First, we start with the virtual cavity model, which results in the Clausius--Mossotti relation. We continue with the real cavity model, where we consider two different configurations. One describes the effective polarisability by treating the particle as a point-like object which leads to the Onsager's real cavity model. In analogy the hard-sphere model corresponds to Onsager's real cavity model and uses spatially spread out dielectric function over the complete cavity volume. This yields the vanishing of the vacuum layer. An interpolation between both real cavity models modeling the finite-size particles is also considered.
\subsection{Clausius--Mossotti relation and virtual cavity model}
The Clausius--Mossotti relation describes the relation between the microscopic quantity polarisability and the macroscopic quantity dielectric function. It is also known as virtual cavity model or local-field corrections, because it describes the increase of the electric field in the presence of a dielectric sphere with radius $R$. By considering the fields inside and outside the sphere ${\bf{E}}'$ and  ${\bf{E}}$ respectively, one finds that the local field around the sphere increases by~\cite{Jackson}
\begin{equation}
 {\bf{E}}' = {\bf{E}}+ \frac{1}{3\varepsilon_0} {\bf{P}} \, , \label{eq:lfc}
\end{equation}
with the polarisation of the sphere ${\bf{P}}$. Using the relation that the polarisation is the electric response of an externally applied electric field
\begin{equation}
 {\bf{P}} = \varepsilon_0\left[\varepsilon(\omega) -1\right] {\bf{E}} \, , \label{eq:polar}
\end{equation}
and that the polarisation is the response to the application of the local electric field to the sphere
\begin{equation}
 {\bf{P}}= \alpha\eta {\bf{E}}'
\end{equation}
with the number density $\eta$ of atoms inside the sphere and the polarisability, one finds the polarisability of a sphere as
\begin{equation}
 \alpha = 4\pi  \varepsilon_0R^3 \frac{\varepsilon(\omega)-1}{\varepsilon(\omega) +2} \, . \label{eq:CMr}
\end{equation}
This equation is the Clausius--Mossotti relation and means that the electric response, expressed by the polarisability $\alpha$, of a sphere with radius $R$ is given by the product of its volume and the Mie reflection.

Furthermore, the local-field correction  due to the presence of a spherical object, Eq.~(\ref{eq:lfc}), together with the polarisation, Eq.~(\ref{eq:polar}), leads us to write the local electric field at the sphere as
\begin{equation}
 {\bf{E}}' = \left(\frac{\varepsilon(\omega)+2}{3}\right) {\bf{E}} \, .
\end{equation}
This results that the selfcorrelation between the local electric field\cite{Buhmann2012}
\begin{eqnarray}
\lefteqn{ \left\langle \hat{\bf{E}}'(\rr,\omega)\otimes \hat{\bf{E}}'^\dagger(\rr',\omega') \right\rangle= \left(\frac{\varepsilon(\omega)+2}{3}\right)^2\left\langle \hat{\bf{E}}(\rr,\omega)\otimes \hat{\bf{E}}^\dagger(\rr',\omega') \right\rangle}\nonumber\\
&&=\left(\frac{\varepsilon(\omega)+2}{3}\right)^2\frac{\hbar}{\pi\varepsilon_0}\frac{\omega^2}{c^2}\mathrm{Im}{\bf{G}}(\rr,\rr',\omega)\delta(\omega-\omega') \, .
\end{eqnarray}
Hence, the local-field corrected excess polarisability can be written as
\begin{equation}
 \alpha^\star(\omega) =  \left(\frac{\varepsilon(\omega)+2}{3}\right)^2\alpha(\omega) \equiv\alpha_{virt}\, , \label{eq:alphaloc}
\end{equation}
which we denote by $\alpha_{virt}$.
\subsection{Onsager's real cavity model}

Onsager's real cavity model considers a spherical vacuum bubble around a particle at position ${\bf{r}}_A$ embedded in a medium with dielectricity $\varepsilon(\omega)$ \cite{Onsager}. Figure~\ref{fig:onsager} illustrates arrangement. The scattering Green function for a spherical two layered system with final and source point in the outer layer can be found in Refs.~\cite{Lewei1994,Tomas2001}. 
The boundary conditions entering the reflection coefficients read as $z = k R_C$ and $z_s=k_s R_C$ with the cavity radius $R_C$ and the absolute value of the wave vector inside and outside of the sphere $k_s$ and $k$, respectively. Considering a single scattering event starting outside the cavity towards its centre and back scattering, these processes can be described with the Born series expansion \cite{Buhmann2012b}. The scattering Green function can be expressed as \cite{Sambale07}
\begin{equation}
 {\bf{G}}^{(S)}_2 (\rr_A,\rr_A,\omega) =R(\omega) {\bf{G}}^{(S)}_{bulk}(\rr_A,\rr_A,\omega) \,,
\end{equation}
 where ${\bf{G}}^{(S)}_{bulk}$ denotes the scattering Green function for a bulk medium with the dielectricity of the outer layer and the transmission coefficient \cite{Sambale07,Abramowitz1972}
\begin{equation}
 \sqrt{R(\omega)} = \frac{i}{z_s\left[j_1(z_s)\left[zh_1^{(1)}(z)\right]'-\varepsilon(\omega)\left[z_sj_1(z_s)\right]'h_1^{(1)}(z)\right]} \, ,
\end{equation}
with the spherical Bessel and first kind Hankel functions $j_n$ and $h_n^{(1)}$, respectively, 
which yields the exact excess polarisability
\begin{equation}
 \alpha^\star(\omega) = -\alpha(\omega)\frac{1}{z_s^2} \left(\frac{1}{j_1(z_s)\left[zh_1^{(1)}(z)\right]'-\varepsilon(\omega)\left[z_sj_1(z_s)\right]'h_1^{(1)}(z)}\right)^2 \, .
\end{equation}
 Applying the Taylor series expansion assuming that the cavity radius $R_C$ is small compared to the relevant wavelengths, $R_C\ll k^{-1},k_s^{-1}$, to this transmission coefficient results
\begin{eqnarray}
 \sqrt{R(\omega)} \approx \frac{3\varepsilon(\omega)}{1+2\varepsilon(\omega)} -\frac{3}{10}\frac{\varepsilon(\omega)\left[10\varepsilon^2(\omega) -9\varepsilon(\omega)-1\right]}{\left[1+2\varepsilon(\omega)\right]^2} \left(\frac{\omega R_C}{c}\right)^2 \, ,
\end{eqnarray}
and leads the famous excess polarisability for small radius $R_C$
\begin{equation}
 \alpha^\star(\omega) = \alpha(\omega) \left(\frac{3\varepsilon(\omega)}{1+2\varepsilon(\omega)}\right)^2 \equiv\alpha_{Ons}\, , \label{eq:alphaOns}
\end{equation}
which will be denoted by $\alpha_{Ons}$. For small, but finite radii, a Taylor expansion
\begin{equation}
 \alpha^\star(\omega) = \alpha(\omega) \frac{3\varepsilon(\omega)}{1+2\varepsilon(\omega)}\left[\frac{3\varepsilon(\omega)}{1+2\varepsilon(\omega)} - 2 \left(\frac{3}{10}\frac{\varepsilon(\omega)\left[10\varepsilon^2(\omega)-9\varepsilon(\omega)-1\right]}{\left[1+2\varepsilon(\omega)\right]^2}\right)\left(\frac{\omega R_C}{c}\right)^2\right] \, ,
\end{equation}
shows that this approximation works for typical cavity radii. It has an agreement of more than 99\% compared with the exact solution. The corresponding frequency of the cavity mode can be estimated as $\omega_c = c/R_C$, with the speed of light $c$. Typical values for the cavity radius are in the order of several {\AA}ngstr\"om which result a cavity mode in the order of $10^{18} \, \mathrm{rad
\, s^{-1}}$, where typical materials are transparent. 
\subsection{Onsager's real cavity model for finite size particles}

The Onsager's real cavity model for spatially extended particles can be described by a spherical three layer system \cite{Sambale10}. The inner layer represents the particle, the outer layer the surrounding medium and the second layer between both denotes the vacuum cavity. The description of this arrangement is similar to Onsager's real cavity model. Again, a scattering process at the outer boundary has to be described that follows the same scattering Green function for a spherically layered system \cite{Tomas2001} whereas the reflection coefficients need to be exchanged by the one for a three layered system which read \cite{chew95,Lewei1994}
\begin{equation}
\tilde{r}_{32}= r_{32}+ \frac{t_{23}r_{21}t_{32}}{1-r_{23}r_{21}} \, , \label{eq:Rt32}
\end{equation}
which denotes the multiple reflection coefficient at the second boundary with the transmission and reflection coefficient $t_{ij}$ and $r_{ij}$, respectively, between the $i$-th and $j$-th layer \cite{Sambale07,Lewei1994,chew95}
\begin{eqnarray}
 r_{i,i+1} &=&\frac{\sqrt{\varepsilon_{i+1}\mu_i}H_n^{(1)}(k_{i+1}a)H_n^{(1)}{}'(k_{i}a)-\sqrt{\varepsilon_i\mu_{i+1}}H_n^{(1)}{}'(k_{i+1}a)H_n^{(1)}(k_ia) }{\sqrt{\varepsilon_i\mu_{i+1}}J_n(k_ia)H_n^{(1)}{}'(k_{i+1}a)-\sqrt{\varepsilon_{i+1}\mu_i} H_n^{(1)}(k_{i+1}a)J_n'(k_ia)   } \, ,\\
 t_{i,i+1} &=& \frac{i\varepsilon_{i+1}\sqrt{\frac{\mu_{i+1}}{\varepsilon_i}} }{\sqrt{\varepsilon_i\mu_{i+1}}J_n(k_ia)H_n^{(1)}{}' (k_{i+1}a) -\sqrt{\varepsilon_{i+1}\mu_i}H_n^{(1)}(k_{i+1}a)J_n'(k_ia)  } \, ,\\
 r_{i+1,i} &=&\frac{\sqrt{\varepsilon_{i+1}\mu_i}J_n(k_{i+1}a)J_n'(k_ia)-\sqrt{\varepsilon_i\mu_{i+1}} J_n(k_ia)J_n'(k_{i+1}a) }{\sqrt{\varepsilon_i\mu_{i+1}}J_n(k_ia)H_n^{(1)}{}'(k_{i+1}a)-\sqrt{\varepsilon_{i+1}\mu_i}H_n^{(1)}(k_{i+1}a)J_n'(k_ia) }\,,\\
 t_{i+1,i} &=& \frac{i\varepsilon_i\sqrt{\frac{\mu_i}{\varepsilon_{i+1}}}}{\sqrt{\varepsilon_i\mu_{i+1}} J_n(k_ia)H_n^{(1)}{}'(k_{i+1}a)-\sqrt{\varepsilon_{i+1}\mu_i}H_n^{(1)}(k_{i+1}a)J_n'(k_ia) }\, ,
\end{eqnarray}
with $J_n(x) = xj_n(x)$ and $H_n^{(1)}(x)=xh_n^{(1)}(x)$ and $a$ denotes the position of the boundary between the $i$-th and $i+1$-th layer. Again, the comparison of the exact Green function for the spherically layered system with that for the bulk propagation, Eq.~(\ref{eq:excesspol}), results in the excess polarisability that reads in general
\begin{equation}
 \alpha^\star(\omega) = -i \frac{6\pi\varepsilon_0}{\sqrt{\varepsilon}}\frac{c^3}{\omega^3} \tilde{r}_{32} \, .
\end{equation}
Note that the dielectric function of the inner sphere is connected to the corresponding polarisability via the Clausius--Mossotti relation, Eq.~(\ref{eq:CMr}). Assuming a small radius of the cavity and of the particle, the vector wave functions reduce to the first order ($n=1$), because all other terms vanish for a centred particle. The reflection coefficient for a purely electric field, Eq.~(\ref{eq:Rt32}), can be expand into a series expansion for small radii $R$ and $R_C$
 \begin{eqnarray}
 \tilde{r}_{32} \approx B_1^N = \frac{2i}{3}\left(\sqrt{\varepsilon\mu}\frac{\omega}{c}\right)^3\left[R_C^3\frac{1-\varepsilon}{1+2\varepsilon}+\frac{9\varepsilon R^3(\varepsilon_s-1)/(2\varepsilon+1)}{(\varepsilon_s+2)(2\varepsilon+1)+2(\varepsilon_s-1)(1-\varepsilon)R^3/R_C^3}\right] \, ,
 \end{eqnarray}
where $R$ and $R_C$ denote the radius of the particle and the cavity, respectively. In analogy to the virtual cavity model, it can be considered  the electromagnetic scattering at a sphere with radius $R_s$ and dielectricity $\varepsilon_s$ embedded in a medium $\varepsilon$. Considering the electric fields through a dielectric sphere with $\varepsilon_s$ and radius $R_s$ embedded in a medium $\varepsilon$, one finds an excess polarisability ($R_s\to R_C$)
\begin{equation}
 \alpha^\star = 4\pi\varepsilon_0 \varepsilon R_s^3\frac{\varepsilon_s-\varepsilon}{\varepsilon_s+2\varepsilon} \equiv\alpha_{HS}\, . \label{eq:polsc}
\end{equation}
which is the Mie coefficient and is denoted $\alpha_{HS}$ meaning that the hard-sphere polarisability arises when the vacuum layer vanishes. Using this result, one can define excess polarisabilities in the three layer model. One is the free space polarisability for the sphere surrounded by vacuum
\begin{equation}
 \alpha_s = 4\pi\varepsilon_0 R^3\frac{\varepsilon_s-1}{\varepsilon_s+2} \, , \label{eq:expols}
\end{equation}
and the excess one for the cavity
\begin{equation}
 \alpha^\star_C = 4\pi\varepsilon_0 \varepsilon R_C^3\frac{1-\varepsilon}{1+2\varepsilon} \, . \label{eq:expolc}
\end{equation}
Substituting Eqs.~(\ref{eq:expols}) and (\ref{eq:expolc}) in Eq.~(\ref{eq:polsc}) the excess polarisability for the three layered system simplifies to
\begin{eqnarray}
  \alpha_{S+C}^\star &=& \alpha_C^\star +\alpha_s\left(\frac{3\varepsilon}{2\varepsilon+1}\right)^2\frac{1}{1+\alpha_C^\star\alpha_s /(8\pi^2 \varepsilon_0^2R_C^6\varepsilon)}\nonumber
  \\
  &=&  \alpha_C^\star +\alpha_s^\star\equiv\alpha_{fs}\, ,\label{eq:expol3}
\end{eqnarray}
which we denote by $\alpha_{fs}$ and the dressed polarisability $\alpha_s^\star$. 
The performed approximation agrees with the exact values with a confidence of more than 99\% for small molecules. Due to the multiple reflections occurring here the quality of the series expansion strongly depends on the radii $R$ and $R_C$. It can be imagined that, for larger object, such as fullerene, this approximation might fail because of the increase of the multiple reflection term. Table~\ref{tbl:summ} summarises the different effective polarisability models.

We note the appearance of the prefactor $\varepsilon$ in the effective polarisabilities of the sphere in Eq.~(\ref{eq:polsc}) (or  and cavity in  Eq.~(\ref{eq:expolc})), suggesting that the effective polarisability of a molecule is generally larger in a more polar medium. This contrasts with the effective polarisability presented by Netz \cite{Netz2004}, which does not contain the prefactor  $\varepsilon$. This apparent contradiction arises because of alternative possible definitions of what the effective polarisability means. The effective polarisability may be thought of as the linear relationship between an external electric field and the dipole (embedded in a medium) induced in a molecule by that field \cite{Jackson}. The polarisation field  ${\bf{P}}$, thought of as the field generated by that induced dipole, has the same value regardless of the definition of the induced dipole. In the picture we have presented here, we have defined the effective polarisability such that the polarisation field corresponds to an induced dipole located inside the medium with  $\varepsilon$.  Netz, by contrast, adopted a definition of the effective polarisability that generates the same polarisation field \emph{as if} the corresponding  induced dipole was located in vacuum, rather than in the medium. In other words, the definition depends on whether we take the effective polarisability to refer to an effective induced dipole in medium or in vacuum.  The difference matters in the evaluation of the van der Waals energy, whether the latter treats the medium explicitly (cf. the appearance of $\varepsilon$ in the  van der Waals parameters  Eq.~(\ref{eq:C3}) or  Eq.~(\ref{eq:C6}) in Sect.~\ref{sect:vdWparameters} below) or whether, as in  Netz's approach, the van der Waals energy is evaluated as  if the field is in vacuum. In our case (effective induced dipole in medium), $\varepsilon$ in the effective polarisability cancels with that in denominator of the van der Waals parameters.  In Netz's case (effective induced dipole in vacuum), the factors of  $\varepsilon$ are never present.  The net van der Waals energy is the same either way (the two definitions were incorrectly mixed in \cite{ParsonsNinham2010dynpol} resulting in an underestimate of van der Waals parameters). We suggest that the approach of effective-dipole-in-medium  provides a more natural or more general definition of the effective polarisability, since in the three-layered system  $\varepsilon$ cannot be simply lifted out from Eq.~(\ref{eq:expol3}). 

\begin{table}
 \caption{Summary of the polarisability models}
 \begin{tabular}{l|l}
  Model & $\alpha^\star = $\\\hline
  Free space & $\alpha$\\
  Local-field corrected, Eq.~(\ref{eq:alphaloc}) & $\displaystyle\left(\frac{\varepsilon+2}{3}\right)^2\alpha $\\
  Onsager, Eq.~(\ref{eq:alphaOns}) & $\displaystyle\left(\frac{3\varepsilon}{1+2\varepsilon}\right)^2\alpha $ \\
  Hard-sphere, Eq.~(\ref{eq:polsc}) & $\displaystyle4\pi\varepsilon_0 R_s^3\frac{\varepsilon_s-\varepsilon}{\varepsilon_s+2\varepsilon}$\\
  Finite-size, Eq.~(\ref{eq:expol3}) & $\displaystyle\alpha_C^\star +\alpha\left(\frac{3\varepsilon}{1+2\varepsilon}\right)^2\frac{1}{1+\alpha_C^\star\alpha /(8\pi^2 \varepsilon_0^2R_C^6\varepsilon)} $
 \end{tabular}
\label{tbl:summ}
\end{table}

\section{Cavity radius for gas molecules in water}

\begin{table}

  \caption{Cavity volume $V_C$ (in general not spherical), cavity radii $R_C$ (spherical approximation), and hard-sphere radii $R$ of four greenhouse gases CH$_4$, CO$_2$, N$_2$O, O$_3$, and other atmospheric gas molecules .}

  \begin{tabular}{l|l|l|l}
    gas & $V_C $ (\AA$^3$)  & $R_C$ (\AA) & $R$ (\AA) \\ \hline
    \ce{CH$_4$}    &  47.00  & 2.239   &  1.6552 \\ 
    \ce{CO$_2$} &   62.28     &   2.459 & 1.7230 \\ 
    \ce{N$_2$O}       & 58.83     & 2.413 &  1.9912\\ 
    \ce{O$_3$}       & 58.79    & 2.412 & 1.9842\\ 
    \ce{O$_2$}       &  43.84     &  2.187 & 1.3003\\ 
    \ce{N$_2$}       & 44.98    &  2.206 & 1.4094\\
    \ce{CO}       &  51.31     &  2.305 & 1.9658\\  
    \ce{NO$_2$}       &  58.33     & 2.406 & 1.9859\\ 
    \ce{H$_2$S}       &  52.73     &  2.326 & 2.0298 \\
    \end{tabular}
  
  \label{table_staticpol_radius}
\end{table}
In order to apply the model of Eq.~(\ref{eq:expol3}) we need to estimate $R$ and $R_C$ for the molecules under investigation. 
Table \ref{table_staticpol_radius} lists the cavity volumes $V_C$ and
corresponding radii $R_C=(3V_C/4\pi)^{1/3}$
that were obtained from the cavity definition
in the recently developed continuum solvent model
implemented in the GPAW package\cite{Enkovaara10,Held14}.
The generally non-spherical cavity with a smooth boundary is obtained from an
effective repulsive potential that describes the interaction between
the continuum of the water solvent and the solute molecule.
This potential leads to an effective solvent distribution
function $0\le g\le 1$ that can be related to measurable partial molar volumes
at the limit of infinite dilution
through the compressibility equation\cite{Kirkwood51}.
Using this connection to fit
the effective potential, only a single parameter is needed to predict
the volumina of various test molecules in water in good
agreement to experiment\cite{Held14}.
The cavity volume $V_C$ is obtained by an integration over the
solvent excluded
volume $1 - g$.

The hard sphere radii $R$ describe the effective spherical radius for the volume occupied by the electron cloud of the molecules \cite{ParsonsNinham2009}, defined as the volume within which the electron density exceeds 0.001 electrons/Bohr$^3$ (evaluated using Gaussian \cite{Gaussian09D01}). The cavity radius $R_C$, by contrast, corresponds to the position at which the electron density of the solvent molecules becomes significant (i.e. the position in space where the dielectric medium starts to respond to an external field) \cite{DuignanNinhamParsons2013solvation,DuignanParsonsNinham2014VolumeEntropy}.

\section{Applications: effective polarisabilities for gas molecules in water}
\label{sect:vdWparameters}

\begin{figure}%
\centering
\subfigure[$\mathrm{N_2O}$]{ \includegraphics[width=0.4\textwidth]{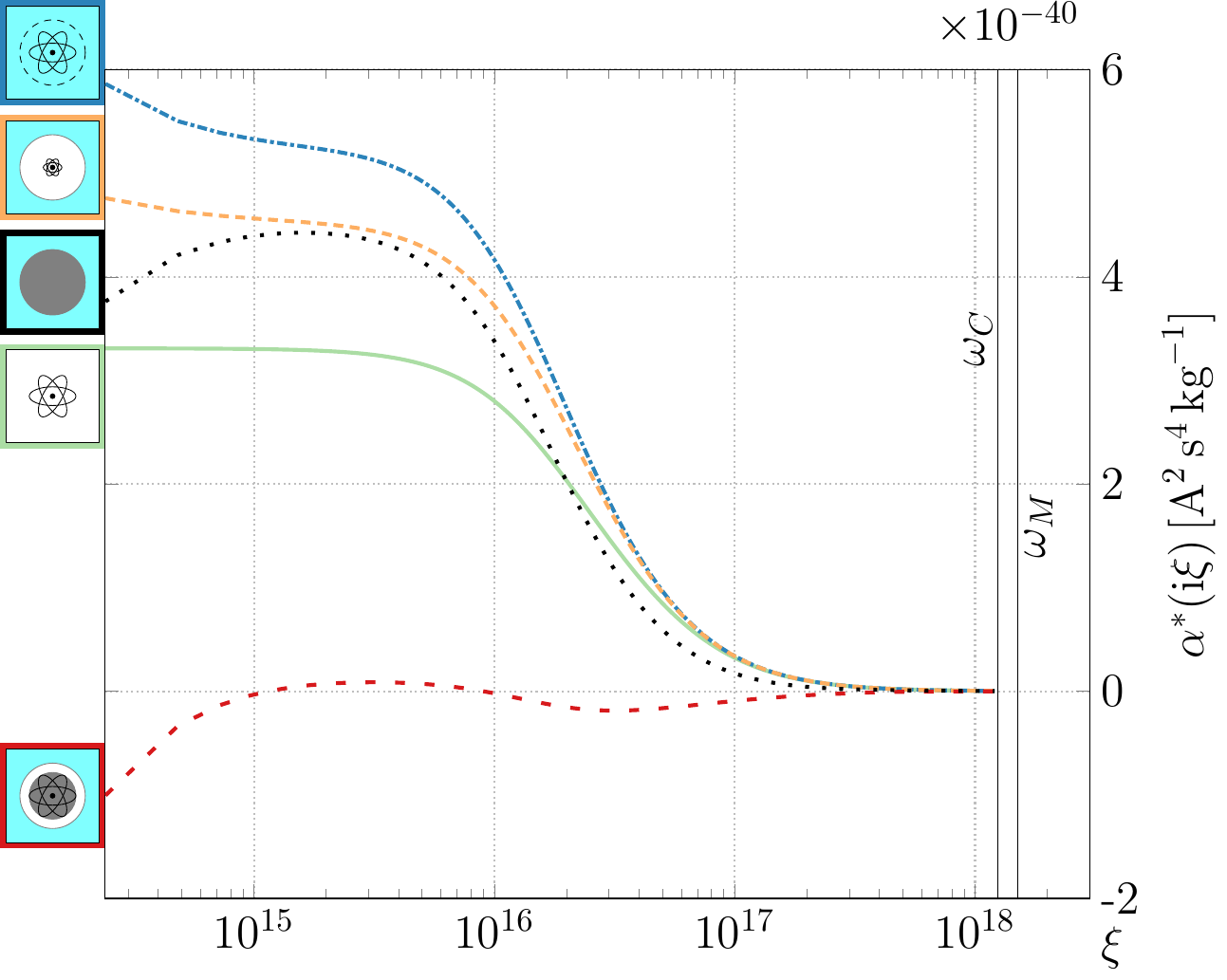}}\qquad
\subfigure[$\mathrm{H_2S}$]{ \includegraphics[width=0.4\textwidth]{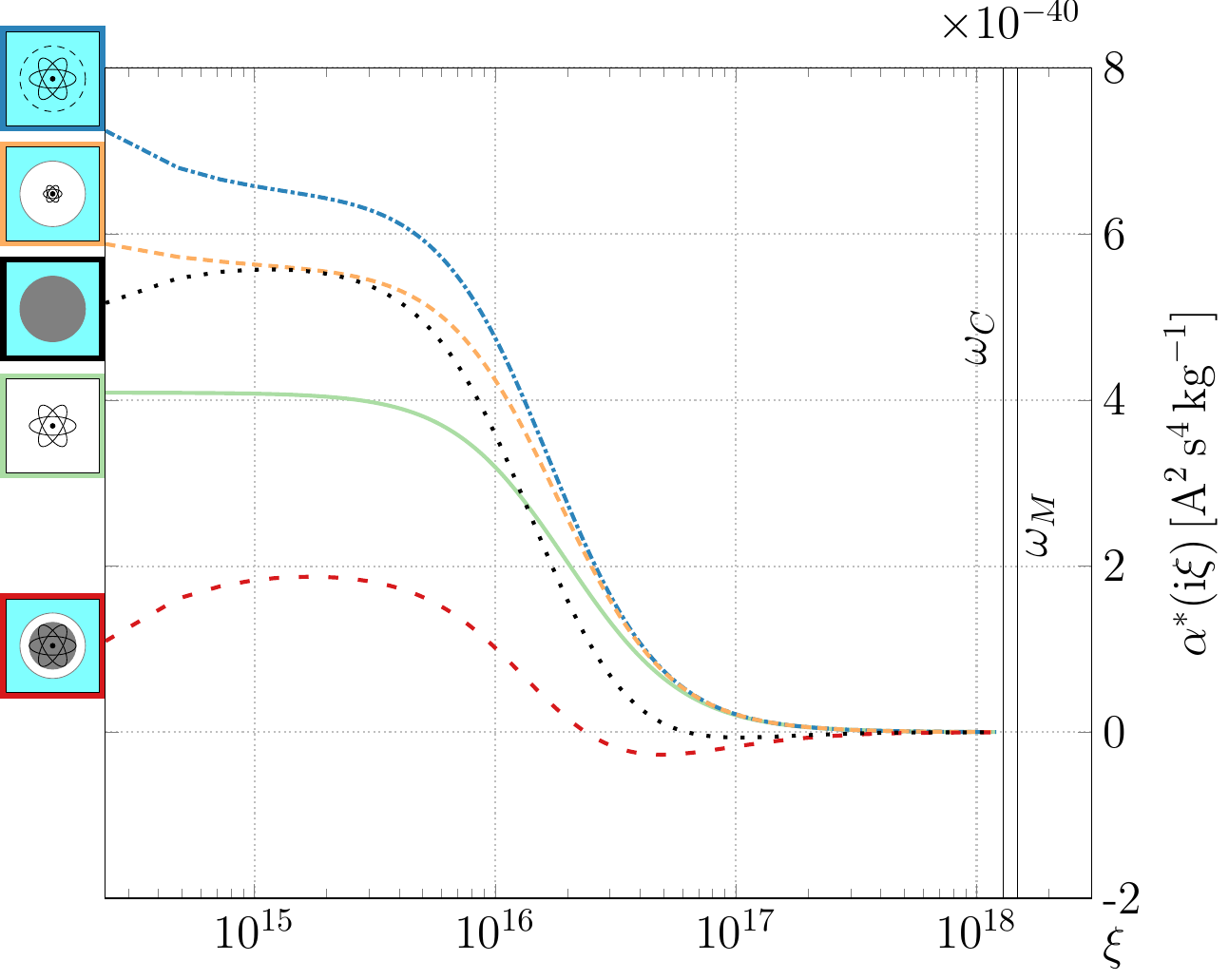}}
\caption{Polarisabilities (green) and effective polarisabilities for (a) $\mathrm{N_2O}$, (b) 
$\mathrm{H_2S}$. 
We compare the local-field corrected model, following Eq.~(\ref{eq:alphaloc}), $\alpha_{virt}$ (blue), the Onsager real cavity model, following Eq.~(\ref{eq:alphaOns}), $\alpha_{Ons}$ (orange), the hard-sphere model $\alpha_{HS}$, Eq.~(\ref{eq:polsc}), (black) and Onsager's real cavity model for finite size particles, Eq.~(\ref{eq:expol3}), $\alpha_{fs}$ (red). In addition, the corresponding cavity modes $\omega_C = c/R_C$ and the hard-sphere modes with respect to the molecule radius $\omega_M = c/R$ are drawn as straight lines.}
\label{3figs}
\end{figure}

The effective polarisabilities determine the strength, and even sign, for van der Waals potentials and Casimir-Polder potentials of gas molecules dissolved in water  near surfaces or in bulk water. The free space polarisabilities and the resulting effective polarisabilities for two different dissolved gas molecules (using the models described in previous sections) are shown in Figure~\ref{3figs}. It is the expectation that the effective polarisability of a molecule in a medium is less than its corresponding free space polarisability since it reflects the difference between the dielectric function in a sphere and the surrounding media (water). This is not  observed in the simpler models $\alpha_{virt}$, Eq.~(\ref{eq:alphaloc}) and $\alpha_{Ons}$, Eq.~(\ref{eq:alphaOns}). The most sophisticated model $\alpha_{fs}$, Eq.~(\ref{eq:expol3}), accounts for a small polarisable sphere (with a size modelled with the hard sphere radius $R$) inside a vacuum bubble (with radius modelled with the cavity radius $R_C$). Here the presence of a vacuum layer means that the effective polarisability may become negative in some frequency regions. This is obviously so since a simple vacuum bubble is less polarisable than the surrounding water. We observe that the closer the values for the hard sphere radius and cavity radius are the less negative the effective polarisability will be. 
In order to illustrate the impact of the different results, we determine the expected van der Waals and Casimir--Polder parameters for the different molecules. 
%


Note that the Casimir-Polder ($C_3$) parameters in Table 3 correspond to more attraction in free space than any of the $C_3$ parameters using the effective polarisability models in water~\cite{DagastinePrieveWhite2000}. The reason is partly the large repulsive contribution from the zero frequency Matsubara term due to the high value for the dielectric constant of water. 
However, it is not only the zeroth frequency that gives repulsion.  Figure~\ref{3figs} shows the effective polarisabilities for methane and nitrous oxide. Based on the threelayer model two subclasses of molecules can be found: one with a purely negative effective polarisability, where $\mathrm{CH_4}$, $\mathrm{CO_2}$, $\mathrm{O_2}$, $\mathrm{O_3}$, $\mathrm{N_2}$, $\mathrm{CO}$ and $\mathrm{CO_2}$ belong to, and one with also positive contributions N$_2$O and H$_2$S. This is caused by the optical density of water. In Eq.~(\ref{eq:expol3}),  $\alpha_C^*$ is always negative and the dressed polarisability $\alpha_{s}^\star$  can dominate it to generate a positive result. This is a direct consequence of the ratio of the optical densities for the considered materials. One needs to take into account for many frequencies where the polarisability is either positive (attractive) and negative (repulsive).
It turns out that the Onsager's real cavity model for finite size particles even predicts repulsion between a molecule in water and a perfect metal surface. This is in contrast to the other models in water. We suggest  to test  the validity of the different effective polarisability models experimentally. Such experiments can be designed very differently. Based on the $C_3$ coefficients given in Table~\ref{tbl:C3}, a solution of $\mathrm{H_2S}$ can be brought horizontally towards a metal surface. Due to the repulsive force, expected for the finite size model, the molecules stabilise at a certain distance due to the equilibrium of the Casimir--Polder and the gravitational forces. 
The results for the van der Waals ($C_6$) parameters are no less interesting and differ by an order of magnitude between the different models. Interactions between equal molecules and between unequal molecular pairs are given in Tables 4 and 5.

\begin{table}
 \begin{tabular}{c|c|c|c|c|c|c}
 gas & vacuum & medium & loc.-corr. & Onsager & finite size & hard-sphere \\ \hline
$\mathrm{CH_4}$ & 2.69 & 2.01 & 2.56 & 2.36 & -0.78 & 3.07\\
$\mathrm{CO_2}$ & 3.61 & 2.8 & 3.46 & 3.23 & -0.96 & 5.71\\
$\mathrm{N_2O}$ & 3.81 & 2.94 & 3.65 & 3.4 & -0.54 & 2.29\\
$\mathrm{O_3}$ & 3.75 & 2.92 & 3.6 & 3.36 & -0.58 & 2.25\\
$\mathrm{O_2}$ & 2.28 & 1.79 & 2.19 & 2.05 & -0.91 & 7.59\\
$\mathrm{N_2}$ & 2.38 & 1.85 & 2.28 & 2.13 & -0.9 & 5.72\\
$\mathrm{CO}$ & 2.44 & 1.87 & 2.34 & 2.17 & -1.29 & 0.1\\
$\mathrm{NO_2}$ & 3.56 & 2.76 & 3.41 & 3.18 & -0.73 & 1.89\\
$\mathrm{H_2S}$ & 3.35 & 2.45 & 3.21 & 2.91 & -0.57 & 1.05
 \end{tabular}
\caption{$C_3$ coefficients in $10^{-49} \mathrm{J m^3}$ for the different gases in vacuum (Eq.~(\ref{eq:C3}) with $\varepsilon(\omega)=1$ and the free-space polarisability), in medium without a cavity (Eq.~(\ref{eq:C3}) with $\varepsilon(\omega)=\varepsilon_{water}(\omega)$ and the free-space polarisability), with the local-field correction (virtual cavity) using Eq.~(\ref{eq:C3}) with $\varepsilon(\omega)=\varepsilon_{water}(\omega)$, the polarisability follows from Eq.~(\ref{eq:alphaloc}), with Onsager's real cavity  using Eq.~(\ref{eq:C3}) with $\varepsilon(\omega)=\varepsilon_{water}(\omega)$ and the polarisability follows from Eq.~(\ref{eq:alphaOns}) and for the real cavity with finite size of the particles using Eq.~(\ref{eq:C3}) with $\varepsilon(\omega)=\varepsilon_{water}(\omega)$ and the polarisability follows from Eq.~(\ref{eq:expol3}) and with the hard-sphere cavity model, Eq.~(\ref{eq:polsc}).}\label{tbl:C3}
\end{table}

\begin{table}
 \begin{tabular}{c|c|c|c|c|c|c}
 gas & vacuum & medium & loc.-corr. & Onsager & finite size &hard-sphere\\ \hline
$\mathrm{CH_4}$ & 116.71 & 48.95 & 155.66 & 79.04 & 6.35 & 230.33 \\
$\mathrm{CO_2}$ & 161.5 & 71.88 & 205.2 & 112.31 & 12.42 & 452.31\\
$\mathrm{N_2O}$ & 188.44 & 82.58 & 245.01 & 130.07 & 7 & 94.63\\
$\mathrm{O_3}$ & 168.7 & 75.01 & 224.46 & 117.06 & 7.85 & 75.4\\
$\mathrm{O_2}$ & 58.02 & 26.61 & 72.78 & 40.88 & 11.16 & 678.26\\
$\mathrm{N_2}$ & 70.29 & 31.52 & 88.03 & 49.07 & 9.33 & 448.07\\
$\mathrm{CO}$ & 79.39 & 34.47 & 104.2 & 54.56 & 16.33 & 7.84\\
$\mathrm{NO_2}$ & 155.27 & 68.87 & 203.93 & 107.68 & 8.95 & 59.28\\
$\mathrm{H_2S}$ & 215.27 & 85.37 & 311.73 & 141.94 & 15.25 & 92.75
 \end{tabular}
\caption{$C_6$ coefficients in $10^{-79} \mathrm{J m^6}$ for the different gases in vacuum (Eq.~(\ref{eq:C6}) with $\varepsilon(\omega)=1$ and the free-space polarisability), in medium without a cavity (Eq.~(\ref{eq:C6}) with $\varepsilon(\omega)=\varepsilon_{water}(\omega)$ and the free-space polarisability), with the local-field correction (virtual cavity) using Eq.~(\ref{eq:C6}) with $\varepsilon(\omega)=\varepsilon_{water}(\omega)$ and the polarisability follows from Eq.~(\ref{eq:alphaloc}), with Onsager's real cavity  using Eq.~(\ref{eq:C6}) with $\varepsilon(\omega)=\varepsilon_{water}(\omega)$ and the polarisability follows from Eq.~(\ref{eq:alphaOns}), for the real cavity with finite size of the particles using Eq.~(\ref{eq:C6}) with $\varepsilon(\omega)=\varepsilon_{water}(\omega)$ and the polarisability follows from Eq.~(\ref{eq:expol3}) and with the hard-sphere cavity model, Eq.~(\ref{eq:polsc}).}
\end{table}

\begin{table}
 \begin{tabular}{c|ccccccccc}
 & $\mathrm{CH_4}$ & $\mathrm{CO_2}$ & $\mathrm{N_2O}$ & $\mathrm{O_3}$ & $\mathrm{O_2}$ & $\mathrm{N_2}$ & $\mathrm{CO}$ & $\mathrm{NO_2}$ & $\mathrm{H_2S}$\\\hline
 $\mathrm{CH_4}$ & 6.35 & 7.65 & 6.4 & 6.4 & 5.85 & 5.99 & 7.84 & 6.77 & 7.33 \\
 $\mathrm{CO_2}$ &  & 12.4 & 8.28 & 9.51 & 11.3 & 10.6 & 13.7 & 10.4 & 4.39 \\
 $\mathrm{N_2O}$ &  &  & 7 & 7.13 & 6.36 & 6.42 & 8.09 & 7.41 & 6.91 \\
 $\mathrm{O_3}$ &  &  &  & 7.85 & 8.09 & 7.76 & 9.94 & 8.32 & 5.27 \\
 $\mathrm{O_2}$ &  &  &  &  & 11.2 & 10.1 & 13.3 & 9.15 & 0.76 \\
 $\mathrm{N_2}$ &  &  &  &  &  & 9.33 & 12.2 & 8.69 & 2.05 \\
 $\mathrm{CO}$ &  &  &  &  &  &  & 16.3 & 11.2 & 2.83 \\
 $\mathrm{NO_2}$ &  &  &  &  &  &  &  & 8.95 & 4.89 \\
 $\mathrm{H_2S}$ &  &  &  &  &  &  &  &  & 15.2
 \end{tabular}
\caption{$C_6$ coefficients in $10^{-79} \mathrm{J m^6}$ for the different combination of interacting particles using the three layer model, Eq.~(\ref{eq:expol3}).}
\end{table}


\section{Impact of particle size}
\begin{figure}[H]
	\centering
	\includegraphics[width=\textwidth]{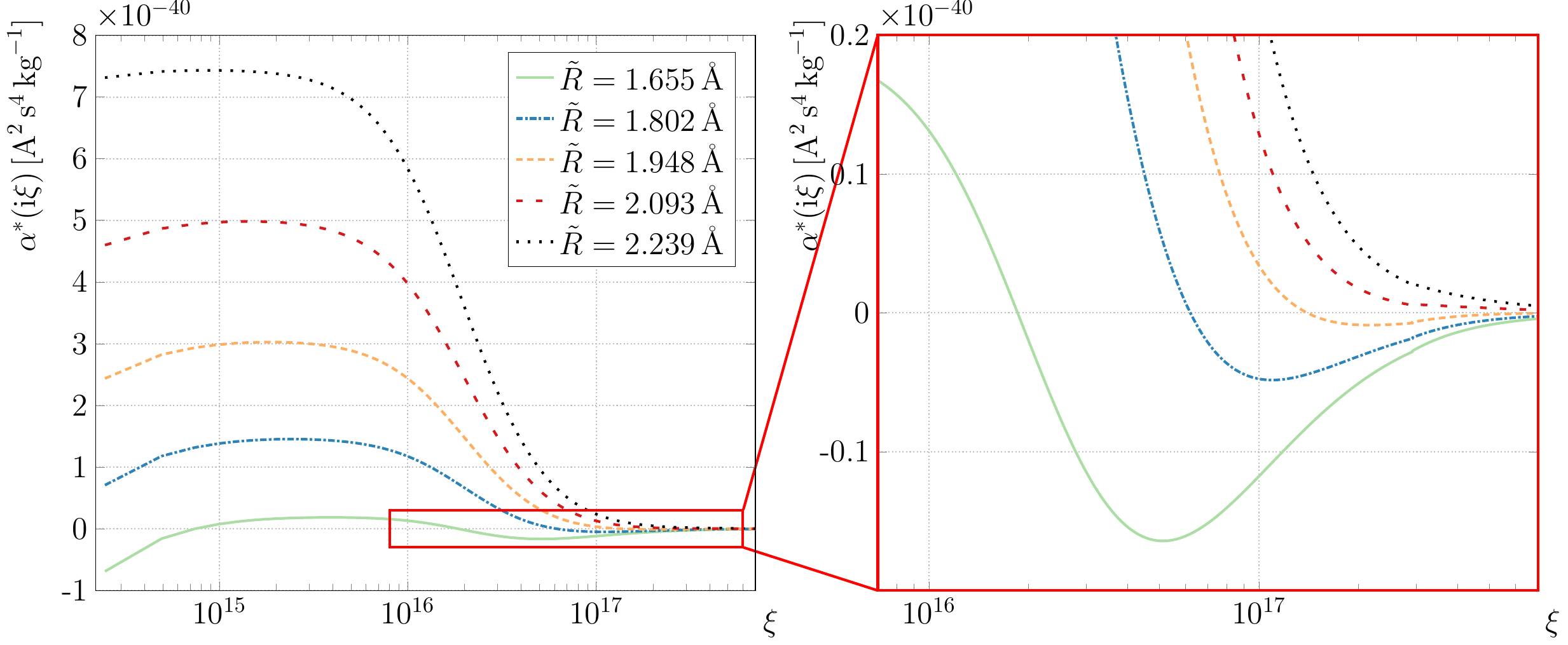}
	\caption{The excess polarisability in the real cavity model for finite size particles (\ref{eq:expol3}) for a toy-methane molecule with the polarisability given by Eq.~(\ref{eq:toymethane}). The radius of the toy-methane sphere is varied between the original radius and the cavity radius, given in Table \ref{table_staticpol_radius}. One sees a transition from partially negative to purely positive polarisabilities.}\label{fig:toymodel}
\end{figure}

The polarisabilities in this model vary strongly between different molecules: while the polarisability for H$_2$S is only negative for high frequencies, the polarisability of CH$_4$ is also negative for the first two Matsubara frequencies. One main difference between H$_2$S and CH$_4$ can be found in Table \ref{table_staticpol_radius}. H$_2$S has a larger sphere radius compared to the cavity radius. Thus, by enlarging the sphere radius of CH$_4$, the polarisability should also become positive for the first two Matsubara frequencies. To verify this, we calculate the leading order excess polarisability (\ref{eq:expol3}) for different sphere radii. The permittivity of the CH$_4$ sphere is fixed using Eq.~(\ref{eq:expols}) with the original sphere radius $R=1.655\,\textup{\AA}$, and the sphere radius $\tilde{R}$ is then varied between the original one and the cavity radius $R_C=2.239\,\textup{\AA}$ (for $R=R_C$, one obtains the hard sphere polarisability (\ref{eq:polsc}) with $R \,\rightarrow \,R_C$). This means we use a radius-dependent polarisability for the methane sphere, given by
\begin{equation}
	\alpha_\text{s}(\tilde{R})=4\pi\varepsilon_0 \tilde{R}^3\frac{\varepsilon_\text{s}(R)-1}{\varepsilon_\text{s}(R)+2}.
	\label{eq:toymethane}
\end{equation}
By varying the radius, one indeed sees a transition to positive excess polarisabilities for small frequencies in Fig.~\ref{fig:toymodel}. Furthermore, the excess polarisability becomes positive for all frequencies for $\tilde{R}\ge 2.033\,\textup{\AA}$.

\section{Conclusions}

We have demonstrated that accurate estimates of the effective polarisability in a medium, and the related Casimir-Polder and van der Waals potentials, may  require a new theory. Such a theory has been worked out in the present paper. However, even more urgent is to test predictions of the real cavity model for finite size particles in a medium. We predict that molecules may be either attracted or pushed away from a metal surface depending on the effective polarisability model. Testing these predictions should pave the way for improved modelling of van der Waals interactions in a medium.

When calculating the effective polarisability of small particles - a subject closely related to Casimir Polder and van der Waals potentials - the presence of a dielectric environment implies complications that are at present not very well understood. In the present paper we have approached this problem from a phenomenological viewpoint, making use of various parameter-based models. In essence they consist of (1) the virtual cavity model, implying the Clausius-Mossotti relation; (2) the real cavity model, which in itself can be divided into two subclasses, the first treating the particle as a point-like object surrounded by a spherical cavity volume which denotes the original Onsager cavity model, the second being a finite-size model of the particle which is now assumed to be surrounded by an annular cavity volume and is an extension of the Onsager model. 
Its limit when the particle radius reaches the cavity radius is the hard-sphere model. We calculated the interaction in terms of scattering Green functions for a spherical three-layer system.

The Green function technique was applied to several greenhouse molecules, as examples. Effective polarisabilities for gas molecules in water were calculated. To illustrate the sensitivity of the formalism with respect to the input parameters, we analysed also a real cavity model for finite size particles for a toy methane model.

As a striking demonstration of the sensitivity of the formalism, we found that molecules can be attracted to, or be pushed away from, a metal surface depending on which effective polarisability model is used. There is thus an obvious need for testing these parameter-dependent theories against experiments in order to improve the modelling of van der Waals interactions in media.

A decision model cannot be made on this theoretical level. 
We note that only the finite-size and hard-sphere models are able to produce repulsive forces, and such forces have indeed been observed in some cases for the related Casimir force \cite{capasso_measured}. Further, the finite-size model is the most detailed description as it takes into account both the finite particle size and the exclusion volume. Based on the results an experimental improvement is given by measuring the expected van der Waals and Casimir--Polder forces. However, deeper theoretical investigations are required to transform the results to measurable spectra, such as refractive indices or molecular extinctions.

\begin{acknowledgement}
We acknowledge support from the Research Council of Norway (Project 250346). We thank the Australian National Computer Infrastructure (NCI).
S.Y.B. gratefully acknowledges support by the German Research Council (grant BU1803/3-1) and the Freiburg Institute for Advanced Studies.


\end{acknowledgement}


\bibliography{achemso-demo,Refs_MW}

\end{document}